\documentclass[10pt,fleqn]{article}

\usepackage{amsmath}
\usepackage{epsf}
\begin{document}

\noindent {\bf\large 
  Quantum Monte Carlo Study of \\
  Random Antiferromagnetic Heisenberg Chain}\footnote{To be published
  in ``Computer SimulationStudies in Condensed Matter Physics XI,''
  Eds. D. P. Landau and H.-B. Sch\"{u}ttler (Springer Verlag, Heidelberg,
  Berlin, 1998).}

\vspace{1em}

\noindent {\it S. Todo, K. Kato, and H. Takayama}

\vspace{1.8em}

\noindent
Institute for Solid State Physics, University of Tokyo, Tokyo 106-8666, 
Japan

\vspace{1.8em} 
\noindent 
{\bf Abstract.} Effects of randomness on the spin-$\frac{1}{2}$ and 1
antiferromagnetic Heisenberg chains are studied using the quantum
Monte Carlo method with the continuous-time loop algorithm.  We
precisely calculated the uniform susceptibility, string order
parameter, spatial and temporal correlation length, and the dynamical
exponent, and obtained a phase diagram. The generalization of the
continuous-time loop algorithm for the systems with higher-$S$ spins
is also presented.

\vspace{2em}
\noindent{\bf\large 1. Introduction}
\vspace{.5em}

\noindent
Low-energy properties of one-dimensional quantum Heisenberg
antiferromagnets have been of great interest for many years.
Particularly, effects of randomness on ground states with strong
quantum fluctuations, such as the so-called Haldane state, have been
studied extensively in recent years [1-5].

In the present work, we investigated $S=\frac{1}{2}$ and $S=1$ random
antiferromagnetic Heisenberg chains by means of the quantum Monte
Carlo method with the continuous-time loop algorithm~[6-9].  To this
end, we first generalize this algorithm to systems with $S=1$ or
higher.  This quite powerful simulation technique enables us
precise calculations of thermodynamic quantities, such as the uniform
susceptibility, etc.

\vspace{1em}
\noindent{\bf\large 2. Model and Numerical Method}
\vspace{.5em}

\noindent
We consider the spin-$S$ antiferromagnetic Heisenberg chain of $L$
spins with random nearest-neighbor interaction:
\begin{equation}
  \label{eqn:hamiltonian}
  {\cal H} = \sum_{i=1}^L  J_i ({\bf S}_i \cdot {\bf S}_{i+1}) \,,
\end{equation}
where ${\bf S}_i$ is a spin operator at site $i$ and ${\bf S}_i^2=S
(S+1)$.  A periodic boundary condition is applied.  All of $J_i$'s are
taken to be positive (antiferromagnetic).

It is well known that traditional world-line Monte Carlo methods~[6]
suffer from long auto-correlation time at low temperatures.  This
difficulty is solved by the loop algorithm~[7,8].  In addition, one
can completely eliminate the systematic error beforehand, which is due
to the discreteness in the imaginary-time direction, by adopting the
recently proposed continuous-time algorithm~[9].  The continuous-time
loop algorithm, which was originally proposed for a spin-$\frac{1}{2}$
system, can be generalized in a straightforward way to the spin-1 or
higher systems as shown below~[10].

First, we introduce spin-$\frac{1}{2}$ representations of the partition function of the Hamiltonian~(\ref{eqn:hamiltonian}):
\begin{equation}
  \label{eqn:partition}
  Z = \sum_{\{n\}} \ \langle n | \, \exp [ -\beta \tilde{\cal H} ] P \, 
  | n \rangle \,,
\end{equation}
where
\begin{equation}
\tilde{\cal H} = \sum_{i=1}^L \sum_{\mu,\nu=1}^{2S}
\frac{J_i}{4} (\sigma_{i,\mu} \cdot \sigma_{i+1,\nu}) \,.
\end{equation}
Each $S_i$ is now represented as a sum of $2S$ Pauli operators
($\sigma_{i,1}$, $\sigma_{i,2}$, $\cdots$, $\sigma_{i,2S}$) and the
trace in Eq.\,(\ref{eqn:partition}) is taken over the complete basis
$\{n\}$ of the space spanned by the $2SL$ Pauli operators.  The
projection operator $P=\prod_i P_i$ projects out unphysical states,
which do not appear in the original phase space of dimension
$(2S+1)^L$.  After performing the Suzuki-Trotter decomposition, we
obtained the partition function of the (1+1)-dimensional classical
Ising model with four-body weights of $\exp [ -\beta J_i
\sigma_{i,\mu} \sigma_{i+1,\nu}/4m]$ and additional $(4S)$-body
weights of $P_i$ at the boundaries in the imaginary-time direction.

Note that each four-body weight is completely equivalent to that
appears in the world-line representation of the spin-$\frac{1}{2}$
system.  Therefore, in constructing a loop algorithm, assignment
probabilities of graphs for these weights remain unchanged.  For a
boundary $(4S)$-body weight, we consider $(2S)!$ types of graphs, each
of which consists of $2S$ edges connecting one of $\frac{1}{2}$ spins
at $\tau=\beta$ with a $\frac{1}{2}$ spin at $\tau=0$ one by one.  A
graph is called `{\em compatible},' if every pair of spins connected
by an edge have a same direction.  A graph is selected out of such
compatible graphs with equal probability.  After the assignment
procedure, loops are identified to be flipped independently with
probability $\frac{1}{2}$.

The continuous-time limit (so-called Trotter limit) of the above
procedure clearly exists.  This is the most significant advantage of
the present algorithm when one makes a comparison with the discrete
version of the general-$S$ loop algorithm, proposed by Kawashima and
Gubernatis~[12].  The present method also works with the XXZ model,
systems in higher dimensions, and even systems with mixed magnitude of
spins without any changes on the algorithm.

We have calculated the spin-spin correlation function of the clean
antiferromagnetic Heisenberg chain with $S=\frac{1}{2}$, 1, $\cdots$,
$\frac{5}{2}$.  The correlation length in the zero temperature limit
is estimated as $\xi\simeq6.02$ and 49.6 for $S=1$ and 2 cases,
respectively, while we observed power-law divergence with the inverse
temperature in the cases with half-integer spins.  The estimates for
$S=1$ and 2 agree quite well with the recent results by the density
matrix renormalization group method [13,14].

\begin{figure}[t]
  \begin{center}
    \leavevmode \epsfxsize=0.66\textwidth \epsfbox{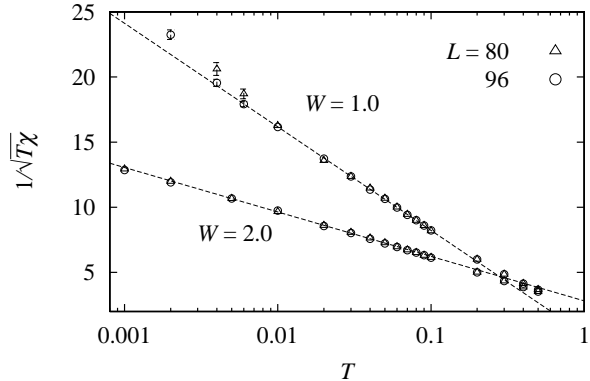}
    \vspace*{-1.5em}
    \caption{Uniform susceptibility in the RS phase of 
      the $S=\frac{1}{2}$ random antiferromagnetic Heisenberg chain
      with bond distribution $(W,u)$=(1,0) and (2,0).
      }
    \label{fig:usus-RS}
  \end{center}
\end{figure}


\vspace{1em}
\noindent{\bf\large 3. Results}
\vspace{.5em}

\noindent
We consider the uniformly-distributed bond randomness with dimerization:
\begin{equation}
  J_i = 1 + (-1)^i u + \delta_i
\end{equation}
with
\begin{equation}
  P(\delta_i) =
  \begin{cases}
    1/W & \text{for $-\frac{W}{2} < \delta_i < \frac{W}{2}$} \\
    0 & \text{otherwise} \,.
  \end{cases}
\end{equation}
The parameters $W$ and $u$ control the strength of the randomness and
the dimerization, respectively.  For $|u|+\frac{W}{2} \le 1$, all
$J_i$'s are antiferromagnetic.

\vspace{1em}
\noindent{\bf 3.1. $\bf S=\frac{1}{2}$ random Heisenberg chain}
\vspace{.5em}

\noindent
The clean (i.e. $W=0$ and $u=0$) $S=\frac{1}{2}$ antiferromagnetic
Heisenberg chain has no excitation gap.  By introduction of
infinitesimal randomness, the system is driven to the random singlet
(RS) phase.  In this phase, it is shown that there is also no
excitation gap, but the antiferromagnetic correlation of spins decays
with a different exponent from that in the clean case~[1,2].

In Fig.\,\ref{fig:usus-RS}, we show the uniform susceptibility in the
cases with $(W,u)=(1,0)$ and (2,0).  The random average was taken
over 3000 samples for the lowest temperature ($T=0.001$) in the
calculation.  The uniform susceptibility diverges as the temperature
goes down, which makes a sharp contrast with the clean ($W=0$) case.
As seen in Fig.\,\ref{fig:usus-RS}, $(\chi T)^{-1/2}$ plotted versus
$\log T$ lies on a straight line at low temperatures in both cases.
This implies the uniform susceptibility behaves as $\chi \sim 1/T(\log
T)^2$, which is consistent with the prediction of the real-space
renormalization group (RSRG) theory~[2].

\begin{figure}[t]
  \begin{center}
    \leavevmode
    \epsfxsize=0.66\textwidth \epsfbox{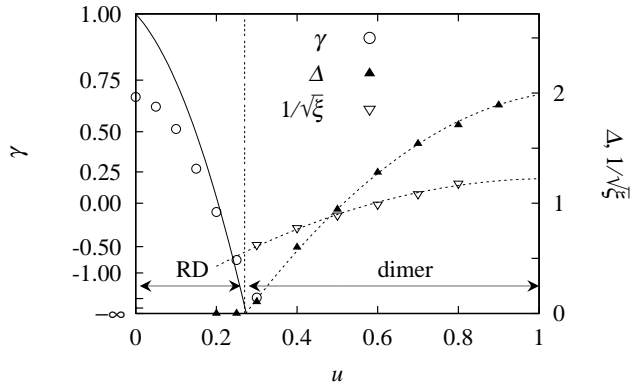}
    \vspace*{-1.5em}
    \caption{Exponent $\gamma$, spin gap $\Delta$, and 
      square root of the inverse spatial correlation length
      $\xi^{-1/2}$ of the $S=\frac{1}{2}$ dimerized random
      antiferromagnetic Heisenberg chain with $u+\frac{W}{2}=1$.}
    \label{fig:phase-1}
  \end{center}
\end{figure}


\vspace{.5em}
\noindent{\bf 3.2. $\bf S=\frac{1}{2}$ random chain with dimerization}
\vspace{.5em}

\noindent
If a weak dimerization $u$ is enforced with $u+\frac{W}{2}$ fixed to 1,
the spin correlation becomes short ranged immediately, while the spin
gap remains closed until $u\simeq0.27$ (Fig.~\ref{fig:phase-1}).  This
phase is referred to as the random dimer (RD) phase, and a realization
of {\it the quantum Griffiths phase} [2,3].  In the RD phase, the
exponent $\gamma$ of the uniform susceptibility ($\chi \sim
T^{-\gamma}$) varies continuously and diverges to the negative
infinity at the end of the RD phase.

The string order parameter $O_{\text{str}}$ exists in the dimer phase
and also in the RD phase.  We found near the RS point ($u=0$),
$O_{\text{str}}$ is well scaled by a logarithmic scaling form:
\begin{equation}
  O_{\text{str}} \sim L^{-2\beta/\nu} f(u L^{1/\nu}, L^{1/\nu}/\log T)
\end{equation}
with $\nu=2$ and $\beta=0.382$ as predicted by the RSRG
arguments~[2,3].  This implies the dynamical exponent $z$ is actually
infinite in the RS phase.

\vspace{1em}
\noindent{\bf 3.3. $\bf S=1$ random Heisenberg chain}
\vspace{.5em}

\noindent
Finally, we consider the $S=1$ chain without dimerization.  We found 
almost similar behavior of the uniform susceptibility as in the case
of the $S=\frac{1}{2}$ dimerized case discussed above.  For small
randomness $W$, the system has the so-called Haldane gap.  As the
strength of the randomness increases, the exponential behavior of the
susceptibility changes to the power-law behavior at $W\sim 1.5$.

For the extremely wide bond distribution ($W=2$), we have confirmed
that the uniform susceptibility obeys the same asymptotic form $\chi
\sim 1/T(\log T)^2$ as in the RS phase of the $S=\frac{1}{2}$ case.


\vspace{1em}
\noindent{\bf\large 4. Summary and Discussions}
\vspace{.5em}

\noindent
Results of the precise quantum Monte Carlo simulation of the random
antiferromagnetic Heisenberg chain have been presented.  We have
obtained the phase boundary between the RD and the dimer phases on
$u+\frac{W}{2}=1$.  We expect there would be a phase boundary {\em
  line} connecting $(W,u)=(0,0)$ and (1.46,0.27) in the $W$--$u$
plane.

For $|u|<\frac{W}{2}$, the bond distributions of odd and even bonds
overlap with each other.  Therefore, it is naturally expected that one
could find a very long and almost {\it clean} segment with arbitrary
small excitation gap.  However, the present result clearly shows that
the spin gap can survive even in such regions.  Its proper
interpretation remains as an open question.

For the $S=1$ chain, similar behavior of the uniform susceptibility
as in the case of $S=\frac{1}{2}$ dimerized chain has been found.
Detailed analysis (critical strength of randomness, etc.) for this
system is now being proceeded and will be presented elsewhere.

\vspace{1em}
\noindent{\bf Acknowledgment}
\vspace{.5em}

\noindent
Numerical calculations in the present work have been mainly performed
using Hitachi SR-2201 at the Supercomputer Center of the University of
Tokyo and DEC cluster workstations and Intel Paragon at the
Supercomputer Center, Institute for Solid State Physics, University of
Tokyo.

\vspace{1em}
\noindent{\bf References}
\vspace{.5em}

\noindent [1] C. Dasgupta and S. K. Ma, Phys. Rev. B {\bf 22}, 1305 (1979).

\noindent [2] D. S. Fisher, Phys. Rev. B {\bf 50}, 3799 (1994); 
B {\bf 51}, 6411 (1995).

\noindent [3] R. A. Hyman, K. Yang, R. N. Bhatt, and S. M. Girvin,
Phys. Rev. Lett. {\bf 76} 839 (1996).  R. A. Hyman and K. Yang, Phys.
Rev. Lett. {\bf 78} 1783 (1997).

\noindent [4] K. Hida, J. Phys. Soc. Jpn. {\bf 65}, 895 (1996); {\bf 65}, 3412 (1996); {\bf 66}, 3237 (1997).

\noindent [5] A. P. Young and H. Rieger, Phys. Rev. B {\bf 53}, 8486 (1996).
A.P. Young, preprint (cond-mat/9707060).

\noindent [6] M. Suzuki, {\it Quantum Monte Carlo Methods in Condensed
  Matter Physics} (World Scientific, Singapore, 1994).

\noindent [7] H. G. Evertz, G. Lana, M. Marcu, Phys. Rev. Lett. 
{\bf 70}, 875 (1993).

\noindent [8] U. J. Wiese and H. P. Ying, Z. Phys. B {\bf 93}, 147 (1994).

\noindent [9] B. B. Beard and U. J. Wiese, Phys. Rev. Lett. {\bf 77}, 
5131 (1996).

\noindent [10] The continuous-time loop algorithm for the $S=1$ 
antiferromagnetic Heisenberg model has been proposed independently in
Ref.~11, which is included in the present algorithm as a special case.

\noindent [11] K. Harada, M. Troyer, and N. Kawashima, preprint 
(cond-mat/97122292).

\noindent [12] N. Kawashima and J. E. Gubernatis, Phys. Rev. Lett. {\bf 73}, 1295 (1994).

\noindent [13] S. R. White, Phys. Rev. Lett. {\bf 69}, 2863 (1992).

\noindent [14] U. Schollw{\"o}ck and T. Jolic{\oe}ur, Europhys. Lett. {\bf 30}, 493 (1995).

\end{document}